\documentclass[twocolumn]{aastex61}
\pdfoutput=1 
\usepackage{amsmath,amstext}
\usepackage[T1]{fontenc}
\usepackage{apjfonts} 
\usepackage[figure,figure*]{hypcap}
\usepackage{gensymb}
\usepackage{natbib}
\usepackage{graphicx}
\usepackage[caption=false]{subfig}

\shorttitle{W49A Draft}
\shortauthors{De Pree et al.}

\begin{document}

\title{FLUX DENSITY VARIATIONS AT 3.6 CM IN THE MASSIVE STAR-FORMING REGION W49A}




 
\author{C. G. De Pree}
\affil{Department of Physics \& Astronomy, Agnes Scott College, 141 E. College Avenue, Decatur, GA, 30030, United States,}
\author{R. Galv\'an-Madrid}
\affil{Instituto de Radioastronom\'ia y Astrof\'isica (IRyA), UNAM, Morelia, Michoac\'an 58089, Mexico,}
\author{W. M. Goss}
\affil{National Radio Astronomy Observatory, Socorro, NM, 87801, United States,}
\author{R. S. Klessen}
\affil{Heidelberg University, Center for Astronomy, Institute for Theoretical Astrophysics, D-69120 Heidelberg, Germany,}
\affil{Heidelberg University, Interdisciplinary Center for Scientific Computing, D-69120 Heidelberg, Germany,}
\author{ M.-M. Mac Low}
\affil{Department of Astrophysics, American Museum of Natural History, New York, NY, 10024, United States,}
\author{T. Peters}
\affil{Max-Planck-Institut f\"{u}r Astrophysik, D-85748 Garching, Germany,}
\author{D. Wilner}
\affil{Smithsonian Astrophysical Observatory, Cambridge, MA, 02138, United States}
\author{J. Bates}
\affil{Department of Physics \& Astronomy, Agnes Scott College, 141 E. College Avenue, Decatur, GA, 30030, United States,}
\author{T. Melo}
\affil{Department of Physics \& Astronomy, Agnes Scott College, 141 E. College Avenue, Decatur, GA, 30030, United States,}
\author{B. Presler-Marshall}
\affil{Department of Physics \& Astronomy, Agnes Scott College, 141 E. College Avenue, Decatur, GA, 30030, United States,}
\author{R. Webb-Forgus}
\affil{Department of Physics \& Astronomy, Agnes Scott College, 141 E. College Avenue, Decatur, GA, 30030, United States,}

\begin{abstract}
A number of ultracompact H~{\sc ii} regions in Galactic star forming 
environments have been observed to vary significantly in radio flux density 
on timescales of 10--20 years. Theory predicted that such variations should 
occur when the accretion flow that feeds a young massive star becomes 
unstable and clumpy. 
We have targeted the massive star-forming region W49A with the Karl G. Jansky 
Very Large Array (VLA) for observations at 3.6 cm with the B-configuration 
at $\sim0\farcs8$ resolution, to compare to nearly identical observations 
taken almost 21 years earlier (February 2015 and August 1994).  
Most of the sources in the crowded field of ultracompact and 
hypercompact H~{\sc ii} regions exhibit no significant changes over this 
time period. However, one source, W49A/G2, decreased by 20\% in peak 
intensity (from 71$\pm$4 mJy/beam to 57$\pm$3 mJy/beam), and 40\% in 
integrated flux (from 0.109$\pm$0.011 Jy to 0.067$\pm$0.007 Jy), where we cite 5$\sigma$ errors
in peak intensity, and 10\% errors in integrated flux. We present 
the radio images of the W49A region at the two epochs, the difference image 
that indicates the location of the flux density decrease, and discuss 
explanations for the flux density decrease near the position of W49A/G2. 
\end{abstract}

\section{INTRODUCTION}
W49 is a radio source consisting of the high mass star forming region W49A and 
the supernova remnant W49B, located $D = 11.11^{+0.79}_{-0.69}$ kpc away 
\citep{zhang2013}. 
The star forming region W49A is one of the largest and most active sites of 
high mass star formation in the Galaxy \citep{sto2014}, and one of the richest clusters known \citep{smith2009}.
W49A contains $\sim$45 H~{\sc ii} regions spread across 
a few  parsecs, of which 13 are classified
as ultracompact (UC) 
H~{\sc ii} regions ($r<$0.05 pc , see \cite{dep1997}).
Recent studies have made attempts to quantify the YSO population (e.g. \cite{saral2015},
\cite{eden2018}), and mapped the molecular gas in the region \citep{gm2013}.
The large number of UC H~{\sc ii} regions distributed over a small region of 
space, as well as the presence of high velocity water maser outflows, makes 
W49A ideal for studying massive star formation and the early evolution of 
H~{\sc ii} regions. The radio appearance of W49A is made up of two well-defined peaks, referred 
to as W49 North (W49N) and W49 South (W49S). In W49N, there are 
$\sim$10 UC H~{\sc ii} regions, each associated with one or more O stars, 
and arranged in a ring 2 pc in diameter \citep{wel1987}. Source W49A/A has a 
bipolar morphology, which suggests ionized outflow \citep{dep1997}. Source 
W49A/G, which is in fact made up of at least 5 smaller components, contains 
the most luminous 
H$_2$O maser outflow in the Galaxy \citep{wal1982}. 

The lifetime of an UC H~{\sc ii} region is estimated to be about 10$^5$ years, 
based on head counts of UC H~{\sc ii} regions in the Galactic disk 
\citep{wood89}. 
This lifetime is approximately 10 times longer than would be expected if UC H~{\sc ii} 
regions simply expand at the thermal sound speed, 
given the number of OB stars in the Galaxy. 
Different theories have been proposed to reconcile this lifetime problem, 
including thermal pressure confinement \citep{dep1995}, turbulent pressure 
confinement \citep{xie1996}, disk photo-evaporation \citep{hol1994}, moving 
star bow shocks \citep{mac1991}, and mass-loaded stellar winds \citep{dys1995}.
None of these models successfully accounts for the observed lifetime of 
UC H~{\sc ii} regions, source morphologies, and a number of other constraints.

A model in which UC H~{\sc ii} region radio emission might vary with time has been 
proposed by \cite{keto2002}, \cite{keto2007} and \cite{peters2010a}. In particular,
\cite{peters2010a} used numerical simulations of star 
formation, which account for heating from both ionizing and non-ionizing 
radiation, to show that the accretion flows that form massive stars become 
gravitationally unstable, resulting in dense clumps and filaments. The fixed 
radiation fields of high mass stars ionize varying volumes of gas, depending on
their density. Therefore, when high-density clumps interacting with stars temporarily
boost the local recombination rate, the H~{\sc ii} region surrounding the star 
can shrink. As a result, the H~{\sc ii} regions associated with a star that is 
still accreting material can flicker between UC and hypercompact (HC) states, 
rather than monotonically expanding as previously assumed. This model potentially 
solves the lifetime problem, as the accretion process continues for 
$\sim$10 times longer than the free expansion time. Independent observations 
of radio recombination lines (RRLs) associated with hypercompact and 
UC H~{\sc ii} regions indicate that some of the most compact H~{\sc ii} regions
continue to accrete material after the H~{\sc ii} region has formed 
\citep{klaasen2017}.

In detail, the model proposed by \cite{peters2010a} predicted that 
UC H~{\sc ii} regions can experience a change of $\sim$5\% yr$^{-1}$ in 
flux density. These variations in flux density can be observed on 
the timescales of available archival observations from powerful radio interferometers.
\cite{gm2011} 
used the models from \cite{peters2010a} 
to predict a $\sim$15\% probability for an increase in flux density, and a 
$\sim$6\% probability for a decrease in flux density over a 20 year time span, 
with statistical uncertainties of 2--3\%. There is strong observational 
evidence of such flux density variations of several H~{\sc ii} regions 
obtained serendipitously from analysis of multi-epoch observations, including 
NGC 7538 \citep{her2004}, 
G24.78+0.08 \citep{gm2008},
Orion BN/KL \citep{riv2015} and NGC 6334I \citep{hunter2018}. 

W49A is the second Galactic massive star forming (MSF) region in which the authors have looked for flux density variations in 
regions containing a large numbers of sources in a single field of view.
\cite{dep2014} previously reported detections of flux density variations in the Sgr B2 Main and North 
regions. 
We detected changes at 7 mm in 2 of 41 ultracompact (UC) and hypercompact (HC) H~{\sc ii} regions, or $\sim$5\% of sources. At 1.3 cm, we detected flux density changes in 4 of 41 UC and HC H~{\sc ii} regions, or about $\sim$10\% of sources. These percentages are roughly consistent with the predictions of \cite{gm2011}, which posited that $\sim$10\% of sources should have detectable flux density variations in a period of 10 years.

In this paper, we present new B-configuration Karl G. Jansky Very Large Array 
(VLA)\footnote{The National Radio Astronomy Observatory is a facility of the National
Science Foundation, operated under cooperative agreement by Associated
Universities, Inc.} 3.6 cm images of the W49A star forming region taken in 2015. 
Archival B-configuration VLA 3.6 cm observations of W49A from 1994 have 
allowed us to search for flux density variations in individual UC and 
HC H~{\sc ii} regions between 1994 and 2015. 
In Section 2, we describe the VLA observations. In Section 3 we report the 
detection of a significant flux density decrease in one source, W49A/G2.
In Section 4 we discuss possible explanations for the detected flux density 
variations.  In the current study of W49A, we have detected a flux density change in 1 of 13 
known UC and HC H~{\sc ii} regions, or $\sim$8\% of sources. The detected flux density decrement is most
likely the result of a high density clump interacting with a star in the G2b or G2c sub regions.

\section{OBSERVATIONS AND DATA REDUCTION}
\subsection{Observations}
W49A was observed with the VLA in the B-configuration at 8.309 GHz (3.6 cm) on August 27, 1994 
and February 8, 2015. The 1994 observations totaled 4.3 hours, with 3.6 hours on source, while the 2015 observations totaled 3 hours, with 2.1 hours on source. The remaining time for both observations was split between the flux density, phase and bandpass calibrator sources. The 1994 data were separated into 63 channels, each with a 97.7 kHz bandwidth. The 2015 data were separated into 16 spectral windows (hereafter SPW), with each SPW separated into 128 channels, each channel being 125 kHz wide. Detailed parameters for both observations are given in Table 1. 

\begin{deluxetable*}{lcc}
\tablecaption{Observational Parameters}
\tablewidth{700pt}
\tabletypesize{\scriptsize}
\tablehead{\colhead{Parameter} & \colhead{1994} & \colhead{2015}}
\startdata
Project Code & AD324 & 15A-089 \\
Date & 1994 August 27 & 2015 February 8 \\
Total Observing Time (hr) & 4.3 & 3 \\
Time on source (hr) & 3.6 & 2.1 \\
Number of Antennas & 27 & 27 \\
Rest frequency (GHz) & 8.3094 & 8.3094 \\
LSR central velocity (km s$^{-1}$) & -53 & 8 \\
Total bandwidth(MHz) & 6.125 & 16 \\
Number of channels & 63 & 128 \\
Channel separation (kHz, km s$^{-1}$) & 97.7 & 125 \\
Flux density calibrator & 3C 286 & 3C 286 \\
Phase calibrator & J1925+2106 & J1925+2106 \\
Bandpass calibrator & J0319+4130 (3C 84) & J0319+4130 (3C 84) \\
FWHM of synthesized beam & 0\farcs72$\times$0\farcs65, -63$\degree$ & 0\farcs71$\times$0\farcs62, -62$\degree$ \\
RMS noise (mJy/beam) & 0.78 & 0.62 \\
\enddata
\tablecomments{Both images were convolved to a circular 0\farcs80 beam before image subtraction as described in the text. The offset in LSR central velocity is the result of centering the 1994 observations between the H92$\alpha$ and He92$\alpha$ Radio Recombination Lines.}
\end{deluxetable*}

\subsection{Data Reduction of VLA 1994 Observations}
Multi-configuration observations of the W49A massive star-forming region made with 
the VLA in the B, C, and D-configurations were reported in \citet{dep1997}. 
These 1994 observations were previously imaged using the Astronomical Image 
Processing System (AIPS). In order to eliminate the possibility of any 
differences in the 1994 and 2015 source flux densities due to differences in 
software implementation of imaging and deconvolution algorithms, the B-configuration
data from 1994 were downloaded from the VLA archive and re-calibrated using the
Common Astronomy Software Application (CASA) package.
The line-free continuum data were then imaged with uniform weighting
and self-calibrated, with three 
phase and one phase and amplitude self-calibration to improve the
signal-to-noise ratio in the resulting image. The final image had an rms noise 
of $\sim$0.58 mJy/beam, and a synthesized beam size of 0\farcs72 x 0\farcs65
(FWHM), BPA$=-$63$^{o}$.

\subsection{Data Reduction of VLA 2015 Observations}
The B-configuration 2015 data presented here are part of a larger 
multi-configuration (A, B, C, and D) dataset that will be fully imaged
and analyzed in an upcoming paper. Here, we present the data reduction of only the portion of 
these 2015 observations that were to be compared to the 1994 data. 
From the larger dataset, SPW 11 (centered at 8.309 GHz) from
the B-configuration observations was separated for 
calibration and imaging, as this most closely matched the 1994 observations
with the old VLA correlator. The 2015 data were flagged and calibrated using
standardized CASA-based calibration techniques. 
A line-free continuum data set was generated from the flagged and calibrated 
data, and was imaged, cleaned and self calibrated using three cycles of phase-only self-calibration 
and one cycle of phase and amplitude self-calibration using CASA,
with the same parameters as the reprocessed 1994 data. 
The final image has an rms noise level of $\sim$0.52 mJy/beam, and 
a synthesized beam 0\farcs71 x 0\farcs63 (FWHM), BPA $=-$62$^{o}$.
 
\section{RESULTS}
\subsection{Aligning and Normalizing Images}
Using CASA, the 1994 and 2015 images were aligned, normalized and convolved 
to the same resolution before subtraction. We first regridded the 1994 image 
from B1950 coordinates to J2000 coordinates,
and then regridded to the image size of the 2015 image.
We then convolved the two images to a 0\farcs8 (circular) beam size 
and normalized the two images.
The normalization factor was determined by taking the ratio of 
the flux densities of 8 of the brightest sources in the central W49N region 
from the two epochs. The entire 2015 image was multiplied by the resulting 
factor of 1.062. This normalization factor lies within the expected mutual uncertainties for the absolute flux scales of the two data sets.
We then aligned the images, 
selecting two UC H~{\sc ii} regions in the 2015 image as 
reference points. 
The regridding and alignment processes introduced only small ($\sim$0\farcs01) shifts.
Figure 1a shows the 3.6 cm
radio emission from the W49N region as observed in 2015, overlaid on the SMA-observed surface density 
maps obtained from CO-isotopologue line ratios \citep{gm2013}.
Figures 1b and 1c show the 1994 and 2015 convolved and normalized images of the central region of W49A,
in the region boxed in Fig. 1a. Source names in Fig. 1a and 1b are from \cite{dep1997}.


\begin{figure*}
\centering
\subfloat[]{
  \includegraphics[width=140mm]{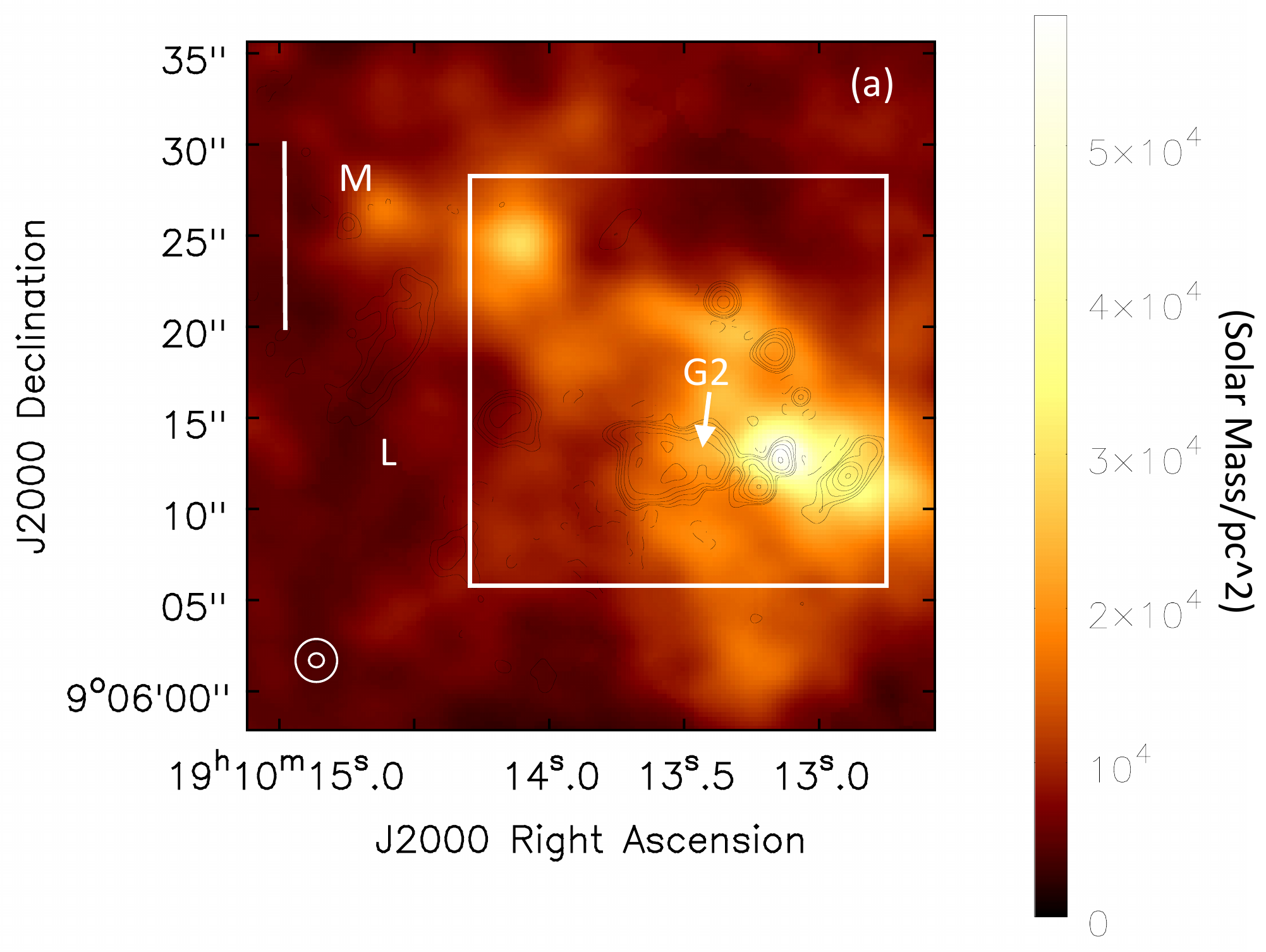}
}
\hspace{0mm}
\subfloat[]{
  \includegraphics[width=90mm]{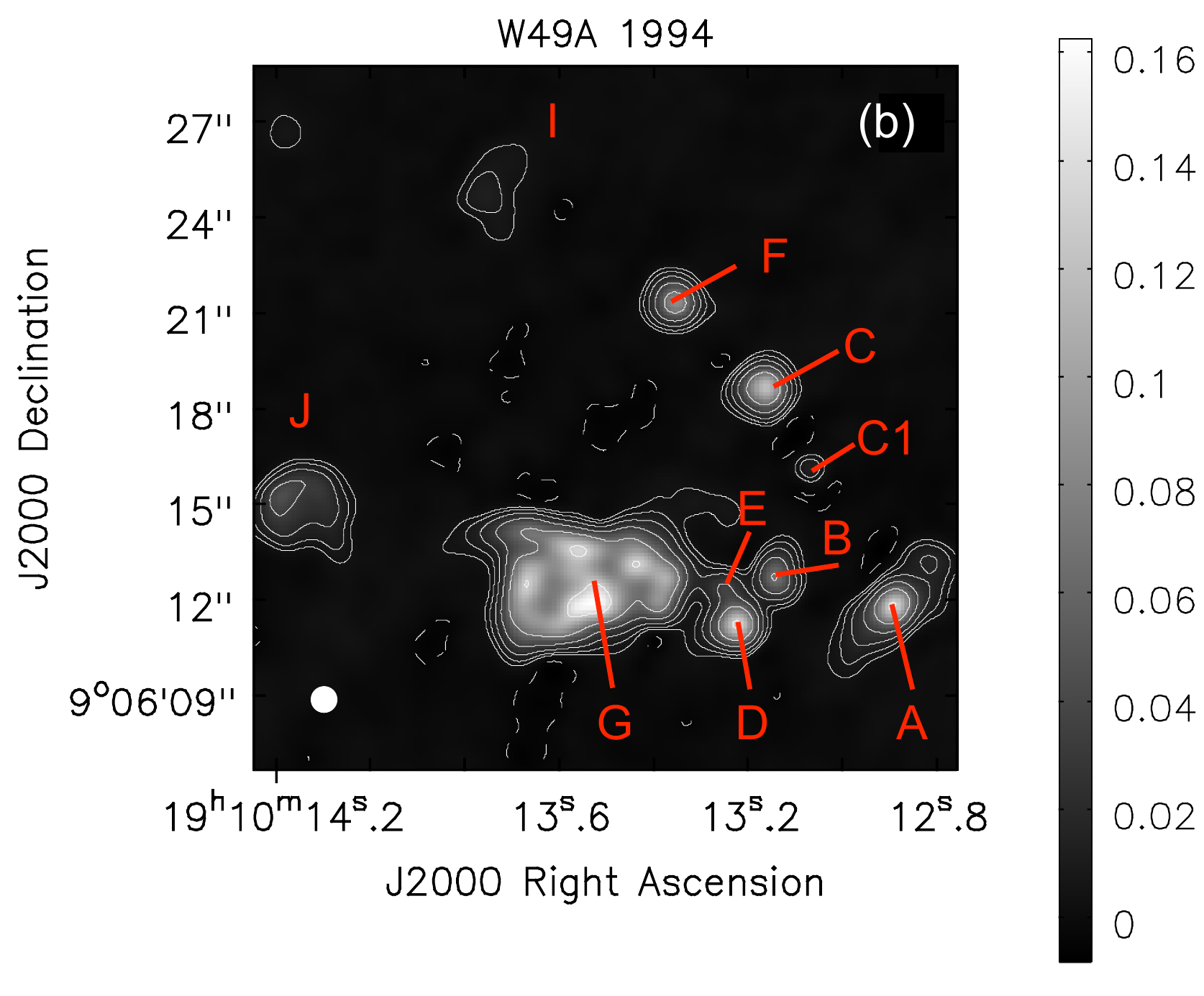}
}
\subfloat[]{
  \includegraphics[width=90mm]{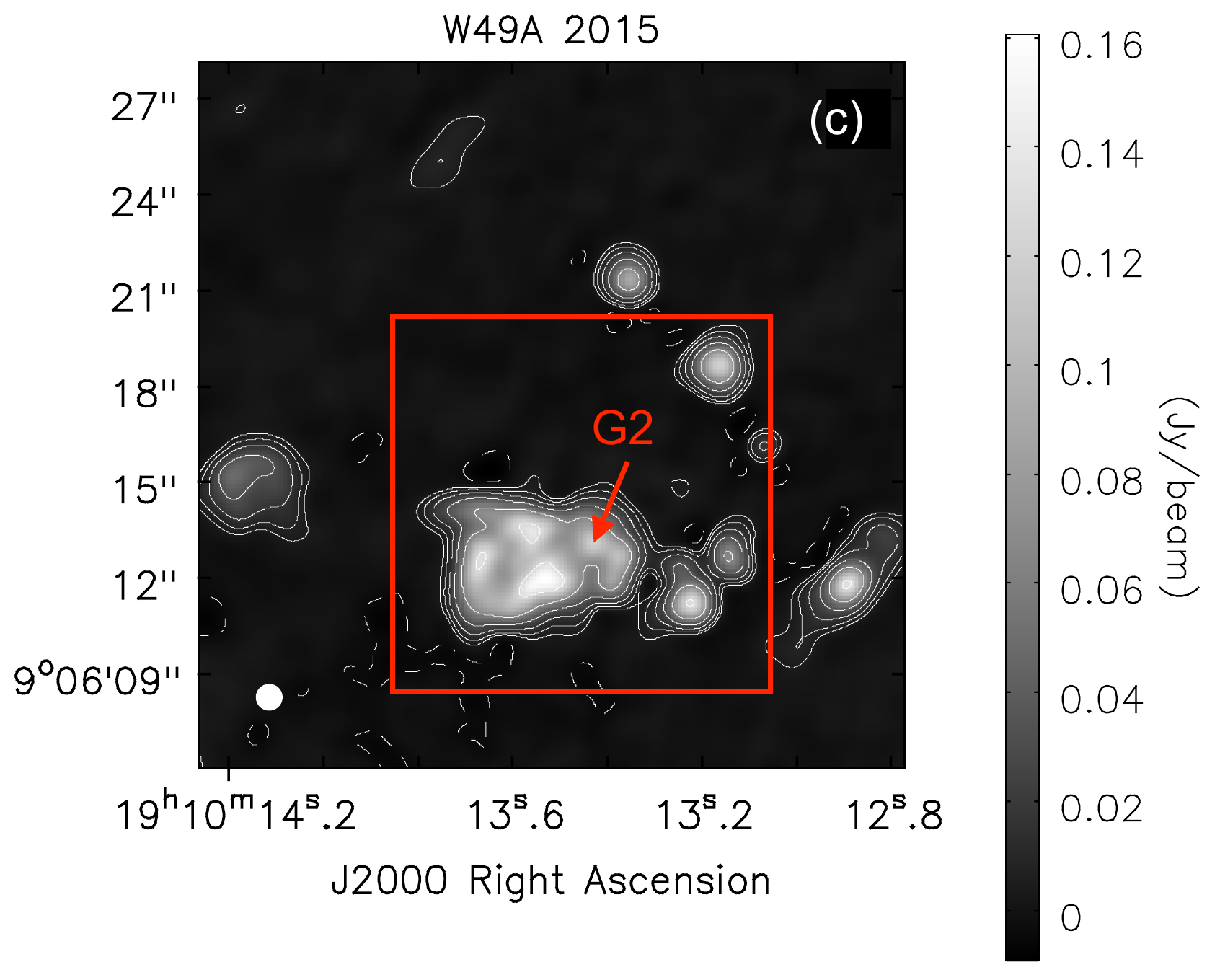}
}
\hspace{0mm}
\caption{W49N as observed in molecular and ionized gas. Figure 1a shows the surface density map 
presented in \cite{gm2013}, with an angular resolution of 2\farcs2 (pseudocolor image) overlaid with the W49A (2015) 3.6 cm continuum contours. Known H~{\sc ii} regions L, M and G2 are indicated in Fig 1a. The box in Fig.\ 1a shows the region of detail in Figures 1b and 1c, which show W49A imaged with the VLA at 3.6 cm in the B configuration in (b) 1994 and (c) 2015. 
Both the (b) and (c) images have a convolved beam size of 0\farcs8x0\farcs8, indicated  by the filled circle in the lower left of each image.
The 1994 observations have an rms noise level of 
$\sim$0.8 mJy/beam, and a peak value of 163 mJy/beam.
First contour for the 1994 observations is at the 5$\sigma$ level  (4 mJy/beam), with other contours at
-1 (dashed), 2, 4, 8, 16 and 32 times the 5$\sigma$ level. 
The 2015 observations have an rms noise level of 
$\sim$0.6 mJy/beam, and a peak value of $\sim$161 mJy/beam.
The 2015 VLA observations shown as contours in (a) and (c) are contoured at the same levels as indicated for (b). The red box in Fig.\ 1b indicates the region of detail shown in Figure 2. The white vertical bar in Fig.\ 1a indicates a size of $\sim$0.5 pc.}
\end{figure*}

\subsection{3.6 cm Difference Images}
Finally, we 
subtracted the 2015 image from the 1994 image to generate a final difference image.
The positional accuracy of the difference image is $\pm$0\farcs1, determined from NRAO-VLA documentation that states typical positional accuracy from standard calibration techniques
is at 10\% of the synthesized beam, or in the case of these data, $\sim$0.1\arcsec.
Figure 2a shows the difference image (contours) superimposed on the convolved 2015 B-configuration 
image (greyscale). The region of this image is shown as a box in Fig. 1c. 
The rms noise in the difference image was measured to be 
$\sigma_{RMS}\sim$1.2 mJy/beam. Dashed contours are plotted at the 
-15$\sigma$, -12.5$\sigma$, and -10$\sigma$ levels, while solid contours 
were plotted at the 10$\sigma$, 12.5$\sigma$, and 15$\sigma$ levels. 
There are two locations in the difference image that exceed the 10$\sigma$ level, and both of those regions $-$ which show a decrease $-$ can be seen in Figure 2a. A single contour ($\sim$10$\sigma$ decrease) is associated with source G5, and a more significant decrease ($\sim$15$\sigma$) on the eastern side of source G2. The first contour was plotted at the 
10$\sigma$ level to ensure that we identified only the most significant regions of flux density change
between the two epochs. In the Discussion below, we consider only the change associated with the G2 region.
\begin{figure*}[t!]
\gridline{\fig{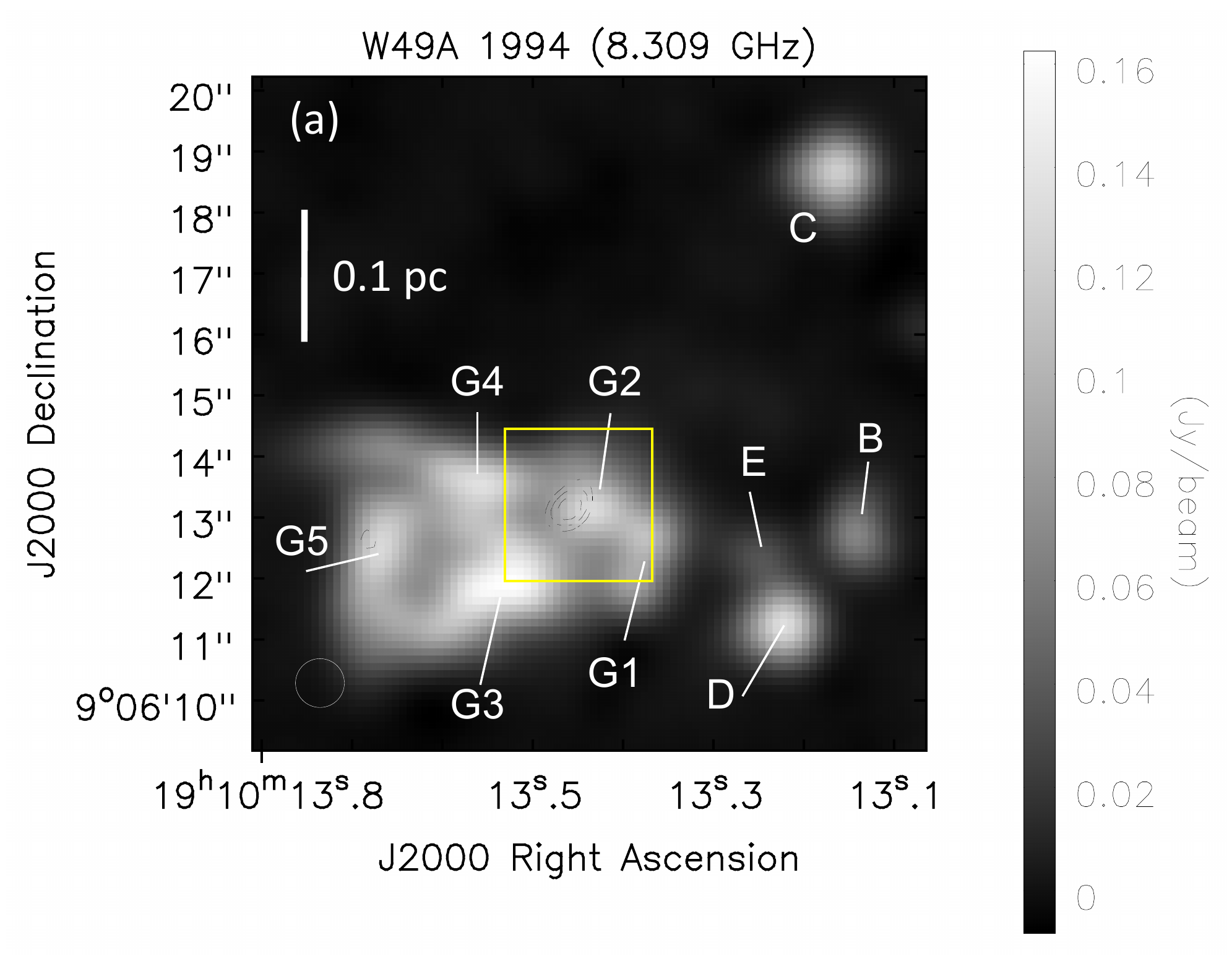}{0.6\textwidth}{}
          }
\gridline{\fig{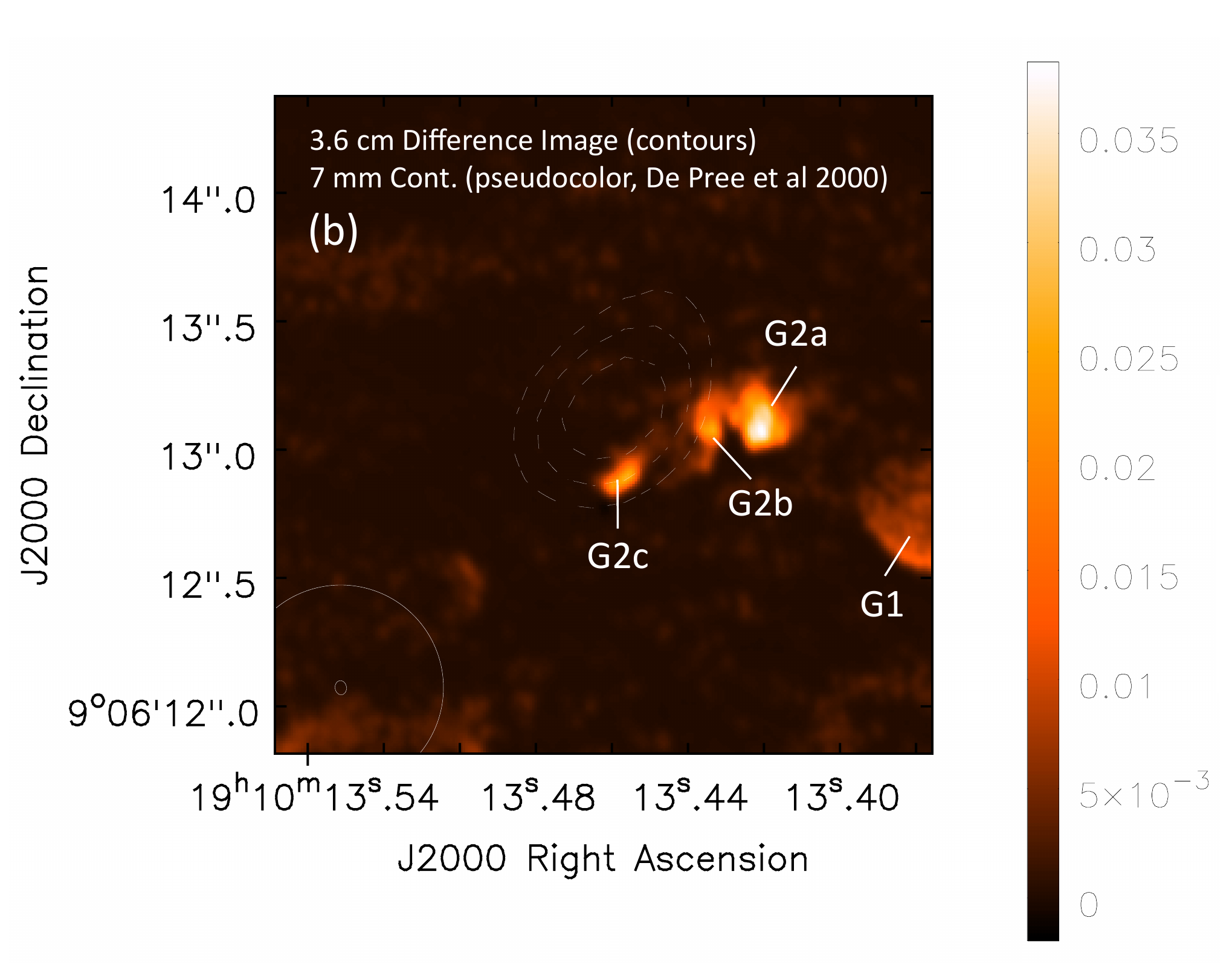}{0.6\textwidth}{}
          }
\caption{(a) The difference image (2015-1994) is shown as black contours
superimposed on the 1994 W49A 8.309 GHz continuum image in greyscale. Contours in the difference image are at -15, -12.5, -10, 10 and 12.5, 15 times the rms noise in the difference image (1.2 mJy/beam). Source G2 is associated with a significant decrease in flux density, greater than 10$\sigma$. There is a single contour at the -10$\sigma$ level associated with the G5 source. Beam size ($\sim$0\farcs8) for the 1994 continuum image and difference image is shown in the lower left hand corner. Sources are named as in \cite{dep1997}. The yellow box indicates the region of detail in Fig. 2b (b) The difference image detail from Fig. 2a is overlaid as contours on the higher resolution VLA 7 mm continuum image \citep{dep2000} in pseudocolor. The synthesized beam size of the two images is indicated by ellipses in the lower left corner. Sub-sources of the G region are indicated as named in \cite{dep2000}.}
\end{figure*}

As an additional check on the flux density change detected in the difference image,
individual sources shown in Fig. 1b were fitted with 2D Gaussians in CASA, 
and Table 2 shows the peak intensity (mJy/beam) and integrated flux (Jy) of 
9 of the 11 UC and hypercompact H~{\sc ii} regions in W49N with peak 
intensities greater than 10$\sigma$ in the image maps. The remaining two sources (G1 and E) are 
not included in this table as the resolution of the B-configuration image 
(0\farcs8) was not sufficient to fit these weaker sources in the crowded 
central W49N region. Errors in Table 2 are 5$\sigma$ in peak intensity and 10\% in
integrated flux. We also tabulate the ratio in peak intensity ({2015}/{1994}) and integrated flux ({2015}/{1994}), with the
errors shown being the errors in the input images added in quadrature.

The result of image differencing shown in Figure 2a 
is confirmed by this sample of source peak intensities and integrated fluxes 
listed in Table 2. Only source G2 shows a significant change in peak intensity,
with a 20\% decrease in peak intensity and 
and a $\sim$40\% decrease in integrated flux. The remaining sources have nearly 
constant values between the two epochs within the uncertainties, and are 
consistent with experiencing no change in flux density between the two 
observations separated by $\sim$21 years.  Thus, in the current study of W49A, we 
have detected a flux density change in 1 of 13 
known UC and HC H~{\sc ii} regions, or $\sim$8\% of sources.

\begin{deluxetable*}{lcccc}
\tablecaption{W49N Source Comparison (1994 to 2015) \\}
\tablewidth{700pt}
\tabletypesize{\scriptsize}
\tablehead{\colhead{Source} & \colhead{Peak Intensity (PI, mJy/beam)} & \colhead{Integrated Flux (IF, Jy)} & \colhead{PI Ratio ({2015}/{1994})} & \colhead{IF Ratio ({2015}/{1994})}}
\startdata
\bf{W49A B} & & \\
1994 & 0.064$\pm$0.004 & 0.102$\pm$0.010 & 1.1$\pm$0.1 & 1.1$\pm$0.1\\
2015 & 0.069$\pm$0.003 & 0.109$\pm$0.011 \\
\bf{W49A C} & & \\
1994 & 0.123$\pm$0.004 & 0.218$\pm$0.022 & 1.0$\pm$0.1 & 1.1$\pm$0.1\\
2015 & 0.122$\pm$0.003 & 0.229$\pm$0.023 \\
\bf{W49A C1} & & \\
1994 & 0.016$\pm$0.004 & 0.019$\pm$0.002 & 1.3$\pm$0.3 & 1.2$\pm$0.2\\
2015 & 0.021$\pm$0.003 & 0.023$\pm$0.002 \\
\bf{W49A D} & & \\
1994 & 0.137$\pm$0.004 & 0.237$\pm$0.024 & 1.0$\pm$0.1 & 1.0$\pm$0.1 \\
2015 & 0.134$\pm$0.003 & 0.231$\pm$0.023 \\
\bf{W49A F} & & \\
1994 & 0.087$\pm$0.004 & 0.132$\pm$0.013 & 1.0$\pm$0.1 & 1.0$\pm$0.1\\
2015 & 0.085$\pm$0.003 & 0.125$\pm$0.013 \\
\bf{W49A G2} & & \\
1994 & 0.071$\pm$0.004 & 0.109$\pm$0.011& 0.80$\pm$0.1 & 0.6$\pm$0.1\\
2015 & 0.057$\pm$0.003 & 0.067$\pm$0.007 \\
\bf{W49A G3} & & \\
1994 & 0.162$\pm$0.004 & 1.17$\pm$0.12 & 1.0$\pm$0.1 & 1.0$\pm$0.1\\
2015 & 0.157$\pm$0.003 & 1.14$\pm$0.11 \\
\bf{W49A G4} & & \\
1994 & 0.142$\pm$0.004 & 0.87$\pm$0.09 & 1.0$\pm$0.1 & 0.9$\pm$0.1\\
2015 & 0.145$\pm$0.003 & 0.80$\pm$0.08 \\
\bf{W49A G5} & & \\
1994 & 0.059$\pm$0.004 & 0.082$\pm$0.008 & 1.1$\pm$0.1 & 1.0$\pm$0.1\\
2015 & 0.062$\pm$0.003 & 0.078$\pm$0.008 \\
\enddata
\end{deluxetable*} 



\section{DISCUSSION}

The G source appears to contain many of the most compact 
H~{\sc ii} regions in W49A \citep{smith2009, dep2000}, suggesting these sources 
are among the youngest and most rapidly accreting in the region, and 
therefore the most likely to undergo radio flux density changes \citep{gm2011}.
The 3.6 cm continuum peak of G2 is also located to the NE of the origin 
of a high velocity outflow traced in water 
masers \citep{smith2009, gw1992}, indicating that this is a highly active, disturbed and young region.

While the B-configuration 3.6 cm data presented here do not have sufficient spatial resolution to 
separate the known sub-sources of G2, we note that higher frequency,
higher spatial resolution (7 mm) observations have indicated the presence of three sub-sources: G2a, G2b and G2c \citep{dep2004, dep2000}. 
Figure 2b shows a detail image of the region around the detected decrease in 3.6 cm flux density in the G2 region (contours),
overlaid with the 7mm continuum emission associated with the G2 sub-sources (pseudocolor). The locations of the nearest 7 mm continuum sources (G1, G2a, G2b and G2c) are indicated in Fig. 2b. 
The region of the detected decrease in 3.6 cm  continuum emission is closest to the 7 mm source labeled G2c, offset by approximately 0.25\arcsec. This offset is larger than the nominal alignment between the 3.6 cm data and the 7 mm data ($\pm$0.1\arcsec), which may have a number of possible explanations. First, the 7 mm data and early epoch 3.6 cm data are not concurrent. The 7 mm observations were made in 1995, 1998 and 2001, while the first epoch of the 3.6 cm data from which the difference image was made were observed in 1993-94. Thus, it is possible that the decrement (observed at 3.6 cm between 1993-94 and 2015) is from a region that had no detectable 7 mm emission by the time of its first high resolution observation in 1998 with the VLA. Second, the two images have very different resolutions (0\farcs8 at 3.6 cm and $\sim$0\farcs05 at 7 mm), and many of the UC sources are optically thick at 3.6 cm, resulting in sources that are centrally brightened at 3.6 cm and edge brightened in the higher resolution (and higher frequency) 7 mm images. Some of the offset thus may be due to the differing resolution and frequency of the two images. Finally, flux decrements interpreted in the context of \cite{peters2010a}, are not expected to occur at the exact peaks of UC H~{\sc ii} regions at all wavelengths, since the decrement position depends on the local position of fast recombination, and the optical depths are different at 3.6 cm and 7 mm. 

Sources G2a and G2b have rising spectral indices (S$_{\nu}\propto\nu^{\alpha}$) between 3.3 mm and 13 mm 
($\alpha\sim$1.1 and 0.8 respectively, \cite{dep2000}), and spectral indices 
with $\alpha\ge$1.0 are predicted for hypercompact H~{\sc ii} regions in 
analytical models \citep{keto2008}, and simulations of gravitationally unstable accretion \citep{peters2010b}.
In addition, G2a and G2b have broad Radio Recombination Lines (RRLs), with $\Delta$V$_{FWHM}$=59$\pm$10
and 58$\pm$14 km/s respectively \citep{dep2004}. While the spectral resolution in these observations is
low (10.3 km/s), both G2a and G2b appear to have multiply-peaked spectra, also indicative of ionized outflow,
and an early evolutionary stage.

Source G2c is the source closest (in projection) to the observed 3.6 cm decrement ($\sim$0\farcs25), and it is is located within the HPBW of the 3.6 cm decrement position. 
Sub-source G2c has an elongated shape
that may indicate the presence of a bipolar ionized outflow from a high mass star in this region, and no detected
RRL emission. 
\cite{smith2009} discuss the detection of a compact mid-IR source (G:IRS1) to the SW of the G2a and G2b
sources. While not coincident with any known UC H~{\sc ii} regions, G:IRS1 is coincident with a hot molecular core.
Indeed, Figure 1a shows that G:IRS1 coincides with an extension in dense gas structures reported in \cite{gm2013} at the center of the W49N `ring' of HC HII regions.
As in many high mass star forming regions, it is clear that star formation activity in W49N is highly clustered, even on scales
of $\sim$0.1 pc. Figure 9 in \cite{smith2009} suggests a model that would place the flux density decrease in the
location of a flattened rotating cloud core, a ready source of material for accretion events.

\cite{peters2010a} suggest that when the accretion 
rate of a protostar suddenly increases due to an increase in the ambient 
density, the surrounding H~{\sc ii} region can contract, and then re-expand again over 
decade to century timescales \citep{gm2011}.  
An accretion event associated with source G2c, or a nearby source no longer observable to its NE, is the most likely source of the
flux density variation detected in the 3.6 cm data between 1994 and 2015.

\section{CONCLUSIONS}
Observations of the W49A region at 3.6 cm at epochs in 1994 and 2015 have 
allowed a careful comparison of H~{\sc ii} region emission over a nearly 
21 year period. One source, G2 (at $\sim$0\farcs8 resolution) was
found to exhibit a significant decrease in both peak intensity (20\%) and 
integrated flux (40\%). This decrease in flux density is consistent with 
a star's radiation being shielded by a high-density clump 
approaching the star in a gravitationally unstable accretion flow,
resulting in an increased recombination rate and reduced volume of ionized gas. 
Source G2c (detected at 7 mm) is possibly associated with this flux density decrease, but there is a $\sim$0\farcs25 offset between the position of the flux density decrease detected at 3.6 cm and the 7 mm position of G2c.

Variable H~{\sc ii} sources like those found in Sgr B2 and W49A should continue 
to be monitored in order to further constrain the model of gravitationally 
unstable accretion \citep{peters2010a}. High resolution molecular imaging 
with the Atacama Large Millimeter/submillimeter Array (ALMA) or the SubMillimeter Array (SMA) of these 
regions will further clarify the interplay between ionizing stars and the 
associated molecular material.

The authors acknowledge thorough and helpful comments from an anonymous referee. CGD and M-MML acknowledge support from NSF RUI award 1615311. RGM acknowledges support from UNAM-PAPIIT program IA102817.

\end{document}